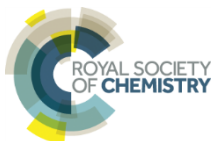

**RSC Advances**

# Delayed hepatic uptake of multi-phosphonic acid poly(ethylene glycol) coated iron oxide measured by real-time Magnetic Resonance Imaging

G. Ramniceanu[1], B.-T. Doan[1], C. Vezignol[2], A. Graillot[3], C. Loubat[3], N. Mignet[1*] and J.-F. Berret[2*]

[1]Unité des Technologies Chimiques et Biologiques pour la Santé (UTCBS), UMR8258/INSERM U1022 CNRS, Chimie ParisTech, 11 rue Pierre et Marie Curie, 75005 Paris, France.
[2]Matière et Systèmes Complexes, UMR 7057 CNRS Université Denis Diderot Paris-VII, Bâtiment Condorcet, 10 rue Alice Domon et Léonie Duquet, 75205 Paris, France
[3]Specific Polymers, ZAC Via Domitia, 150 Avenue des Cocardières, 34160 Castries, France

**Abstract**
We report on the synthesis, characterization, stability and pharmacokinetics of novel iron based contrast agents for magnetic resonance imaging (MRI). Statistical copolymers combining multiple phosphonic acid groups and poly(ethylene glycol) (PEG) were synthesized and used as coating agents for 10 nm iron oxide nanocrystals. *In vitro*, protein corona and stability assays show that phosphonic acid PEG copolymers outperform all other coating types examined, including low molecular weight anionic ligands and polymers. *In vivo*, the particle pharmacokinetics is investigated by monitoring the MRI signal intensity from mouse liver, spleen and arteries as a function of the time, between one minute and seven days after injection. Iron oxide particles coated with multi-phosphonic acid PEG polymers are shown to have a blood circulation lifetime of 250 minutes, *i.e.* 10 to 50 times greater than that of recently published PEGylated probes and benchmarks. The clearance from the liver takes in average 2 to 3 days and is independent of the core size, coating and particle stability. By comparing identical core particles with different coatings, we are able to determine the optimum conditions for stealth MRI probes.

**keywords**: Magnetic resonance imaging - Contrast agents - Iron oxide nanoparticles - PEGylated coating – Pharmacokinetics

## 1 – Introduction

Magnetic Resonance Imaging (MRI) is one of the most widespread non-invasive imaging techniques in clinical practice and research. Current MRI methods are based on the intrinsic contrast of soft tissues and vessels, and provide important information on a broad range of pathologies. Novel imaging techniques have emerged over the last decade, in particular techniques based on the use of contrast agents.[1-3] Most studied MRI contrast agents are gadolinium chelates and magnetic particles. Made from iron oxide, magnetic particles provide several advantages among which a substantial biodegradability over time and an absence of cytotoxicity. In addition, depending on their size the particles can be used as positive or negative contrast agents, leading to an enhancement of longitudinal and transverse proton relaxation rates.[4-6]

For applications, the pharmacokinetic profile of the MRI probes has to be known. Absorption, *in vivo* biodistribution, metabolic transformation and elimination from the tissues and organs are the primary mechanisms involved in pharmacokinetics. In the blood pool, nanoparticles administered intravenously are recognized from plasma proteins, which adsorb at the particle surface spontaneously.[7-12] The protein binding process, known as the *protein corona* formation prevents the particles to interact specifically with potential targets. Simultaneously, protein adsorption activates the particle uptake by the mononuclear phagocytic system through the circulating macrophages and monocytes. The two-step opsonization mechanism described here is responsible for the particle elimination from the blood stream and for their accumulation in unrelated organs, typically the liver and the spleen.[13] As a result, commercial contrast agents using iron oxide cores (e.g. Endorem® from Guerbet, Resovist® and Cliavist® from Bayer) were first designed for imaging hepatic lesions and tumors.[1] Ever since, iron oxide particles have been surveyed in the context of other clinical applications, including heart transplantation, brain lesions imaging and tumor targeting.[2,6,14-16]

To prevent opsonization, studies have shown that neutral hydrosoluble polymers such as poly(ethylene glycol) (PEG) or polysaccharides are efficient.[13,17-20] Tethered at the chain extremity or from several monomeric units, the polymers in aqueous environment form a swollen brush that acts as a protective layer against protein adsorption and particle aggregation. In this context, efforts were mainly directed towards the molecular weight effect on the *in vitro* stability and *in vivo* stealthiness.[15,21-27] From *in vivo* pharmacokinetics studies, Leal *et al.* have found an optimum PEG value of 3 kDa,[25] whereas particle accumulation in tumor was found to be efficient with 10 kDa polymers.[15] Recently, Ruiz *et al.* reported a doubling of the residence time in blood for PEG conjugated iron oxide particles, and a reduction in the liver and spleen uptake.[21,22] In some of the previous studies, the PEGylated contrast agents tested were prepared from 10 nm particles; however it was found that during synthesis or functionalization the particles aggregated and formed large clusters containing tens to hundreds of particles.[15,21,22,26,28] For aggregates, additional parameters such as size and morphology were found to affect the MRI contrast properties and the biodistribution *in vivo*. These properties were also different from those of single nanoparticles.

Another crucial issue in particle functionalization is the nature and strength of the link between the coat and the inorganic surface. A wide variety of techniques based on covalent or non-covalent binding are now achievable, including ligand adsorption, layer-by-layer deposition or surface-initiated polymerization.[29] Non-covalent strategies based on the assembly of separately synthesized components are known to exhibit enhanced yields in terms of quantity of particles produced.[30] Novel macromolecular architectures obtained by radical polymerization were also considered.[31] Na *et al.* designed multi-dentate catechol and PEG derivatized oligomers that provide colloidal stability over a broad range of pH and electrolyte concentrations.[32] Single or multiple phosphonic acid based polymers were also synthesized.[31,33-35] Phosphonate groups are known to have a higher affinity towards metallic atoms or metal oxides compared to sulfonates and carboxylates, and are hence able to build stronger links with surfaces. Sandiford *et al.* exploited PEG polymer conjugates containing a terminal bis-phosphonic acid group for binding to magnetic nanoparticles.[34]

We have recently shown that multi-phosphonic acid PEG-copolymers with up to 4 anchoring groups per chain conferred considerable stabilization to 10 nm iron oxide cores. The coated particles were found to be stable for months in extreme salt and pH conditions, as well as in cell culture media. In a previous work, multi-phosphonic acid PEG coated particles were tested against fibroblast and macrophage cultures and exhibited exceptional low cellular uptake.[12] Here, these newly coated particles are assessed *in vivo* using a live mouse model. MRI is particularly suited for this study as it provides spatial and temporal resolved images with high

# RSC Advances

contrast. The present survey reveals that the phosphonic acid PEG copolymers with multiple anchoring groups are able to prolong the particle lifetime in the blood up to 3 hours, *i.e.* 50 times greater than that of anionic particles or Cliavist® benchmark, and 10 times greater than that of recently published PEGylated probes.[21-23] A qualitative model taking into account the uptake and clearance is proposed and is found to be appropriate for all contrast agents studied. By comparing identical core particles with different coating, we are able to determine the optimum conditions for stealth MRI probes.

# 2 - Materials & Methods

## 2.1 - Iron oxide nanoparticles and Coating

### 2.1.1 - Iron oxide nanoparticles

Iron oxide nanoparticles were synthesized by alkaline co-precipitation of iron(II) and iron(III) salts.[36,37] The salts were dissolved in a hydrochloric solution and co-precipitated by a concentrated base (ammonia, $NH_3$). The particles were then concentrated magnetically and the supernatant was removed. The concentrated phase was washed with water using magnetic sedimentation followed again by supernatant removal. Nitric acid ($HNO_3$) was then added to the particles down to pH 1.5. Addition of a large excess of ferric nitrate at water boiling temperature led to the oxidation of magnetite into stable maghemite nanocrystals. The nanoparticles were sorted according to their size by successive phase separation steps.[38] At pH 1.5, the particles are positively charged and have nitrate counterions adsorbed on their surfaces. The resulting interparticle interactions are repulsive and impart an excellent stability to the dispersion.[37,39] Here, two $\gamma$-$Fe_2O_3$ nanoparticles batches (diameter 6.8 nm and 13.2 nm) were synthesized (Figs. 1a-d). The magnetic and geometric size distributions were obtained from vibrating sample magnetometry (VSM) and from transmission electron microscopy (TEM), respectively. Table I provides a list of the diameters and dispersities (ratio between standard deviation and average diameter) obtained from these two techniques.

|  | 6.8 nm $\gamma$-$Fe_2O_3$ nanoparticle | 13.2 nm $\gamma$-$Fe_2O_3$ nanoparticle |
|---|---|---|
| diameter VSM (nm) | 6.7 | 10.7 |
| dispersity VSM | 0.21 | 0.33 |
| diameter TEM (nm) | 6.8 | 13.2 |
| dispersity TEM | 0.18 | 0.23 |
| molecular weight $M_W^{Part}$ (Da) | $1.3 \times 10^6$ | $12 \times 10^6$ |
| hydrodynamic diameter $D_H$ (nm) | 14 | 27 |

*Table I*: *Characteristics of the iron oxide particles used in this work. The bare particle size and size dispersity were determined by vibrating sample magnetometry (VSM) and by transmission electron microscopy (TEM). The weight-averaged molecular weight $M_W^{Part}$ was obtained from static light scattering.[40] $D_H$ is the bare particle hydrodynamic diameter in water.*

The nanoparticle magnetization was also measured using VSM and its value was found slightly lower than that of bulk maghemite[41,42] ($2.9 \times 10^5$ A $m^{-1}$ instead of $3.9 \times 10^5$ A $m^{-1}$). The particle crystallinity was studied by electron microdiffraction scattering. Five diffraction rings characteristic of the maghemite structure were observed (S1-S4). In the dispersed state, light scattering techniques allowed the measurements of the weight-averaged molecular weight and of the hydrodynamic diameter for the particles.[40] Throughout the paper, the concentrations are expressed in terms of the iron molar concentration, noted $[Fe]$. For instance, the $[Fe]$ = 5 mM dispersion used for the intravenous injection corresponds then to a weight concentration of 400 µg mL$^{-1}$. Assuming a blood pool of 1 mL for wild-type BALB/c mice (age 8 weeks), the concentration of circulating particles is estimated at 40 µg mL$^{-1}$. The contrast agent used as a benchmark, Cliavist® was commercialized by the company Bayer (Germany). According to supplier specifications, the Cliavist® dispersion has an iron concentration of 500 mM and contains 4 nm magnetic core particles aggregated into 55 to 65 nm clusters.[1] The 1400 Da carboxydextran molecules used in the formulation serve both as a linker for the maghemite nanocrystals, the polymer role being indeed to prevent the dispersion sedimentation.

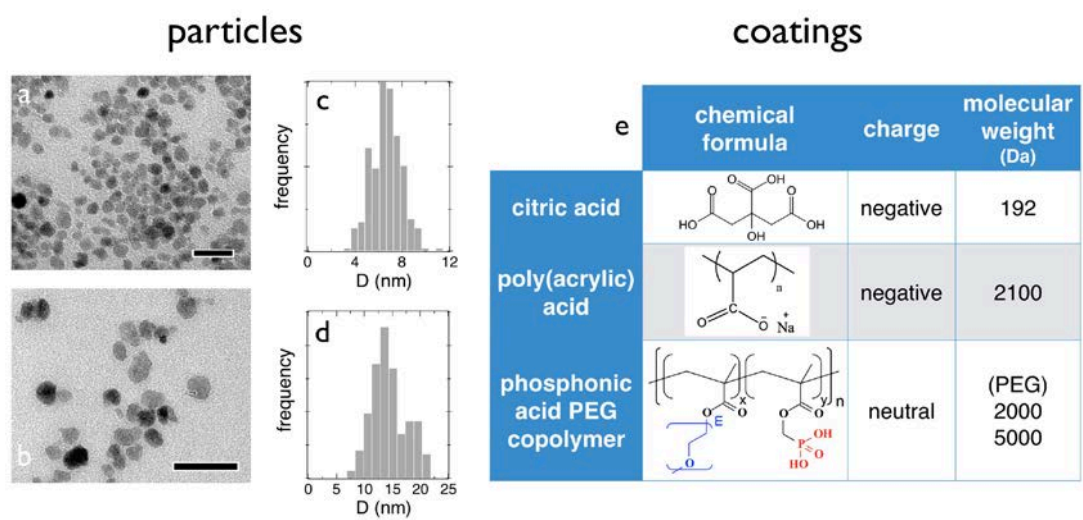

**Figure 1:** *a, b) Transmission electron microscopy (TEM) of the 6.8 and 13.2 nm iron oxide nanoparticles. The bars are 20 nm and 50 nm respectively. c, d) Particle size distribution determined by TEM. e) Table showing the chemical formula, charge and molecular weight of the organic coatings used in this study. For the phosphonic acid PEG copolymer, only the PEG molecular weight is indicated.*

*2.1.2 – Coating*

In this section, we describe the macromolecules and protocols used to coat the iron oxide particles (Fig. 1e). Poly(ethylene glycol methacrylate-*co*-dimethyl(methacryoyloxy) methyl phosphonic acid), abbreviated in this work phosphonic acid PEG copolymer was synthesized by Specific Polymers®, France (*http://www.specificpolymers.fr/*). The synthesis was carried out by free radical polymerization of PEG-methacrylate (PEGMA, SP-43-3-002, CAS: 26915-72-0) and dimethyl(methacryoyloxy)methyl phosphonate (MAPC1, SP-41-003, CAS: 86242-61-7) monomers. Synthesis details can be found in previous reports.[12,43] The phosphonic acid PEG copolymers were characterized from $^1$H NMR and $^{31}$P NMR using a Bruker Avance 300 spectrometer operating at 300 MHz. For the copolymers, a molar-mass dispersity of 1.8 was obtained by size exclusion chromatography on PolyPore column using THF as eluent and polystyrene standards. From the molar equivalent of acid groups obtained from NMR and molecular weight determination, the number of phosphonic acids and PEG segments was estimated. This number was found to be 3.1 ± 0.2 and 3.7 ± 0.2 for copolymers with PEG$_{2K}$ and PEG$_{5K}$, respectively (Table II). Here and below, the error bars are defined as the standard deviations (SD). These determinations confirm the existence of multiple functional groups on the same polymer backbone.

| polymer name | acid group meq/g | functional group per chain | PEG density 6.8 nm | PEG density 13.2 nm |
|---|---|---|---|---|
| copolymer with PEG 2000 Da | 0.873 | 3.1 ± 0.2 | 1.9 | 1.1 |
| copolymer with PEG 5000 Da | 0.349 | 7.4 ± 0.4 | 2.0 | 1.1 |

*Table II: Structural parameters of the PEG copolymers used in this work. The first column denotes the molar equivalent of acid groups per gram ($milli\,eq\,g^{-1}$) of polymer as determined by $^1H$ NMR. The numbers of the functional groups are determined from the molecular weight and from the molar equivalent of acid groups.[12] The error bars are here defined as the standard deviations (SD).*

For the coating, we used a protocol derived in 2008.[44,45] Dispersions of oppositely charged iron oxides and phosphonic acid PEG copolymers at the same concentration ($c$ = 0.1 wt. %) and same $pH$ ($pH$ 2) were mixed at increasing volume ratios $X$ between $10^{-3}$ and $10^3$. The $pH$ of the mixed solution was then raised to 8 by ammonium hydroxide addition, leading to well-dispersed coated particles. The mixed polymer/particle dispersion stability diagram was investigated by dynamic light scattering (Fig. S5). By plotting the hydrodynamic diameter against the mixing ratio $X$, a transition between individual coated particles and micron-sized aggregates occurs at the critical value, $X_C$. The existence of a critical mixing ratio suggests that adsorption occurred *via* a non-stoichiometric electrostatic binding process.[44,45] The model assumes that for each value of $X$, the polymers are equally distributed among the particles in the dispersion.[45] Below $X_C$, the iron oxide surfaces are saturated with polymers, the functional end-groups exceeding the number of binding sites. Above, the coverage is incomplete and the particles precipitate upon $pH$ increase (as uncoated particles do). Here, we exploit this feature to derive the number of adsorbed chains per particle $n_{ads}$:[12]

$$n_{ads} = \frac{1}{X_C}\frac{M_n^{Part}}{M_n^{Pol}} \qquad (1)$$

where $M_n^{Part}$ and $M_n^{Pol}$ are the particle and polymer molecular weights, respectively. For PEG$_{2K}$ on 6.8 nm and 13.2 nm particles, $X_C$ = 1.3 and 5, leading to a number of adsorbed polymers $n_{ads}$ = 97 and 207, respectively. For PEG$_{5K}$, $X_C$ = 0.5 and 2, resulting in $n_{ads}$ = 43 and 88 (Table S5). These $n_{ads}$-values correspond to a PEG density of 1 - 2 nm$^{-2}$. With zeta potentials of -6 mV, electrokinetic measurements confirmed that the multi-phosphonic acid PEG coated particles were globally neutral (S4). Before use, the dispersions were finally dialyzed against deionized water using a 50 kD cut-off membrane to remove the excess polymer, and further concentrated by ultrafiltration. They were then autoclaved (Tuttnauer Steam Sterilizer 2340M) at 120 °C and atmospheric pressure during 20 min to prevent bacterial contamination, and stored at 4 °C in a secure environment.

Citric acid is a weak triacid of molecular weight $M_w$ = 192 Da, with acidity constants pK$_{A1}$ = 3.1, pK$_{A2}$ = 4.8 and pK$_{A3}$ = 6.4. Surface charge complexation with citric acid was performed during the synthesis through simple mixing. At pH 8, citric acid molecules are ionized, and particles are coated with citrate ions. As ligands, citrates are characterized by adsorption isotherms and adsorbed species are in equilibrium with free citrates in the bulk. The free citrate concentration was kept at the value of 8 mM,[39] both in DI-water and in culture medium. The hydrodynamic citrate coated particle diameter was identical to that of bare particles, indicating a layer thickness under 1 nm.

Poly(sodium acrylate), the salt form of poly(acrylic acid) with a weight-average molecular weight $M_w$ = 2.1 kDa and a dispersity of 1.7 was purchased from Sigma Aldrich and used without purification. To adsorb polyelectrolytes on the particles, the precipitation-redispersion protocol was applied.[41] The precipitation of the iron oxide dispersion by $PAA_{2K}$ was performed in acidic conditions (pH2). The precipitate was then separated by magnetic sedimentation and its pH was increased by ammonium hydroxide addition. The precipitate redispersed spontaneously at pH8. $PAA_{2K}$ coated γ-$Fe_2O_3$ hydrodynamic sizes were around 4 – 6 nm larger than the hydrodynamic diameter of the uncoated particles, indicating a corona thickness of 2 – 3 nm. Electrophoretic mobility and zeta-potential values are provided in the Supporting Information section (S4). As for the multi-phosphonic acid PEG coated particles, the dispersions were dialyzed and autoclaved to prevent bacterial contamination.

## 2.2 - Methods

### 2.2.1 - Dynamic light scattering (DLS)

Light scattering experiments were performed on a NanoZS spectrometer (Malvern) at the wavelength of 633 nm and in backscattering configuration (scattering angle 173°). From the scattered intensity evolution, the second-order autocorrelation function of the light is calculated and analyzed using different procedures provided by the instrument software. These procedures are the cumulant method and the CONTIN algorithm. Both gave consistent values for the hydrodynamic diameter $D_H$. As mentioned previously, the nanoparticle hydrodynamic diameters are found to be systematically larger than those obtained by other techniques, in particular by electron microscopy and magnetometry (Table I). The reason for this difference is related to the particle size distribution, along with the fact that light scattering is sensitive to the largest objects of the distribution.

### 2.2.2 - Cell Culture

Adherent cells from mouse hepatocyte cells BWTG3 were studied. Hepatocyte cells BWTG3 were grown in T25-flasks as a monolayer in Dulbecco's Modified Eagle's Medium (DMEM) with high glucose (4.5 g $L^{-1}$) and stable glutamine (PAA Laboratories GmbH, Austria). The medium was supplemented with 10% fetal bovine serum (FBS) and 1% penicillin/streptomycin (PAA Laboratories GmbH, Austria). Exponentially growing cultures were maintained in a humidified atmosphere of 5% $CO_2$ and 95% air at 37°C, and in these conditions the plating efficiency was 70 – 90% and the cell duplication time was 12 – 14 h. Cell cultures were passaged twice weekly using trypsin–EDTA (PAA Laboratories GmbH, Austria) to detach the cells (6 millions in average for T25 flasks) from their culture flasks and wells. The cells were pelleted by centrifugation at 1200 rpm for 5 min. Supernatants were removed and cell pellets were re-suspended in assay medium and counted using a Malassez counting chamber.

### 2.2.3 - Toxicity assay

The method measured the cellular mitochondrial activity. Subconfluent cell cultures (60% confluency at treatment time) on 96 well plates were treated with 100 µL/well of nanoparticles at different concentrations for 24 h, culture medium was removed, cells were rinsed with culture media without phenol red and incubated with 100 µL/well of 2-(4-iodophenyl)-3-(4-nitrophenyl)-5-(2,4-disulphophenyl)-2H-tetrazolium (WST-1, Roche Diagnostics), diluted 1/10 in culture medium without phenol red for 1 to 4 h. The assay is based upon the reduction of the tetrazolium salt WST-1 to formazan by cellular dehydrogenase. The generation of the dark yellow colored formazan was measured at 450 nm in a multiwell-plate reader against a blank containing culture media and WST-1 and it was corrected from the absorbance at 630 nm. The supernatant optical density is directly correlated to cell number.

*2.2.4 – Relaxometry*

Images were acquired with a 7 Tesla spectrometer equipped with a $^1$H radiofrequency linear coil of inner diameter 40 mm (Bruker, Karlsruhe Germany). $T_1$ and $T_2$ relaxation times were measured on 200 µL solutions with increasing iron concentrations, [Fe] = 0.02, 0.05, 0.1, 0.2 and 0.5 mM to determine the longitudinal and transverse relaxivities $r_1$ and $r_2$. For $T_1$-measurement, a saturation sequence with 8 different repetition times from 60 ms to 15 s was used to evaluate the proton spin-lattice recovery. The field of view was of a 3×3×0.15 cm$^3$ with a pixel resolution of 0.234×0.234×1.5 mm$^3$ in all protocols. For $T_2$-measurement, a sequence with 32 sequential echo times from 11 ms to 385 ms was applied to follow the decay of the proton spin-spin relaxation. The total scan time was 30 min. The relaxivities rates $r_{1,2}$ were obtained by adjusting the signal intensity of each agent at different concentrations using:[46] $1/T_{1,2} = r_{1,2}[Fe] + 1/T_{1,2}^0$. These measurements were made in physiological saline solution and in DMEM with serum.

*2.2.5 - In vivo MRI (including data analysis)*

All animal work was performed in accordance with the institutional animal protocol guidelines in place at the University Paris-Descartes (Saisine CEEA34.JS.142.1) and approved by the Institute's Animal Research Committee. Wild-type female 8 weeks BALB/c mice were anaesthetized by isoflurane inhalation (1.5% isoflurane in a 1:1 mixture of air and oxygen at the rate of 0.25 L min$^{-1}$) and placed in a dedicated contention cradle in a 7 Tesla MR-scanner. 100 µL of iron oxide nanoparticles in saline 0.9 wt. % with an optimized concentration of $[Fe]$ = 5 mM were intravenously injected *via* the tail vein thanks to a catheter specifically elaborated for non-magnetic use. The injected iron concentration was adjusted so that the hepatic uptake and clearance processes take a maximum of one week. For reproducibility studies, 2 to 4 mice were injected with the contrast agents tested. 13.2 nm citrate and PEG$_{5K}$ coated particles were injected on one mouse only. For imaging purposes, two types of time resolution were used sequentially: a high temporal resolution with low spatial resolution for the first 180 minutes, and a lower time resolution with high spatial resolution for time range between 20 min and 7 days. With this protocol, it was ensured that the early stage of the hepatic uptake could be evaluated accurately. The low spatial resolution scans were compared and normalized to the higher resolution images using a zero filling technique for signal optimization. All images were recorded with a field of view of 3×3 cm$^2$, corresponding to an in-plane resolution of 125×125 µm$^2$ in both sequences. The scanning protocol was developed using Paravision 5.1 software. To minimize motion artifacts, two methods of respiratory synchronization were used. The high spatial, low temporal resolution sequence synchronized the acquisition with the mouse breathing cycles. The low spatial, high temporal resolution sequence used an internal navigator that detects the mouse respiratory and cardiac cycles and reconstructs the images in a post-processing operation. This second method is independent of the breathing cycle changes. The time frame was set to one minute for the fast scans and three minutes for the slower scan. To recapitulate the chronological imaging experiments, images were recorded before injection and during the first 20 min at a rate of one image per minute, then at 30 min, 40 min, 1 h, 1h40, 2h40, 4h, 7h20 and at 1, 2, 3, 4 5, and 7 days after injection. 3 mice per days were investigated with the addition of 3 to 9 mice in the follow-up process between one day and one week. To study the probe biodistribution, several organs including liver, spleen, kidney and arteries (Supporting Information S6) were monitored. Comparison with commercial Cliavist® was also performed.

# 3 - Results and Discussion
## 3.1 - Iron oxide based contrast agents and coating

Iron oxide nanoparticles of diameters 6.8 nm and 13.2 nm were synthesized by alkaline co-precipitation of iron salts and subsequent oxidation into maghemite ($\gamma$-$Fe_2O_3$) and coated with different macromolecules.[36,37] As coating agent, we used a statistical copolymer that combines multiple phosphonic acid groups and poly(ethylene glycol) chains on the same backbone. The full name of the polymer is poly(ethylene glycol methacrylate-*co*-dimethyl(methacryoyloxy) methyl phosphonic acid), abbreviated phosphonic acid PEG copolymer in the sequel of the paper. Synthesized by free radical polymerization (Specific Polymers®, France), the polymers were designed to strongly adsorb onto metal and metal oxide surfaces via the multiphosphonic acid groups.[12] Copolymers with 2 kDa and 5 kDa PEG chains were synthesized and studied.

The aforementioned PEGylated copolymers were compared to other coating types, including citric acid and poly(acrylic acid) polymer (PAA). Citric acid is a well-known and widely used coating for inorganic particles. In the molecular imaging context, citrate coated particles were tested in a clinical phase I trial to evaluate their pharmacokinetics and safety.[47] Poly(acrylic acid) in contrast was less studied and to our knowledge the biodistribution of PAA coated contrast agents was not reported yet. *In vitro*, it was shown that PAA coat is biocompatible and non-toxic, and that the charged polymer layer brings significant protection against protein adsorption and particle agglomeration.[9,48-50] A recent *in vivo* study has revealed however acute effects on cardiovascular function.[49] The chemical formulae, charge and molecular weights of the coating agents probed are summarized in Fig. 1b. More details on the coating and characterization methods are given in the M&M section.

Important parameters for the particles are the total hydrodynamic diameter $D_H$ and the coating thickness $h$, which were determined by light scattering (Table III). The polymer layers were found to have the same thickness for the 6.8 and for the 13.2 nm particles, and to agree with the brush structure.[51,52] In good solvent conditions, $PEG_{2K}$ is a chain of gyration radius 1.3 nm,[53] whereas its fully stretched length is about 10 nm. $h$-values of 5 ± 1 nm (Table III) are intermediate and indicate that the chains at the surface are slightly stretched, typically by 40 – 50 %. The $PEG_{5K}$ layer on the 6.8 nm iron oxide has a thickness of 16 ± 2 nm, which corresponds to a higher stretched configuration.[52] PAA layers have a thickness of 2 – 3 nm, as mentioned in earlier work.[9,40,44,50]

| nanoparticles | coating | hydrodynamic diameter $D_H$ nm | Layer thickness $h$ nm |
|---|---|---|---|
| $\gamma$-$Fe_2O_3$ 6.8 nm | bare | 14 ± 1 | 0 |
| | citrate | 15 ± 1 | 0.5 ± 1 |
| | $PAA_{2K}$ | 18 ± 1 | 2 ± 1 |
| | $PEG_{2K}$ | 22 ± 1 | 4 ± 1 |
| | $PEG_{5K}$ | 46 ± 3 | 16 ± 2 |
| $\gamma$-$Fe_2O_3$ 13.2 nm | bare | 27 ± 1 | 0 |
| | citrate | 28 ± 1 | 0.5 ± 1 |
| | $PAA_{2K}$ | 33 ± 1 | 3 ± 1 |
| | $PEG_{2K}$ | 37 ± 2 | 5 ± 1 |
|  | PEG$_{5K}$ | 44 ± 3 | 8.5 ± 2 |
| Cliavist© 4 nm | carboxy dextran | 66 ± 4 | aggregates |

***Table III***: *Characteristics of the iron oxide particles used in this work. $D_H$ denotes the particle hydrodynamic diameter and $h$ the coating thickness as determined from dynamic light scattering measurements. Data for Cliavist® are from ref.[1] The error bars denote here the standard deviations.*

### 3.2 - Stability in physiological and culture media

As for the evaluation of protein corona and particle agglomeration, the following protocol was applied.[9,44] A few microliters of a concentrated dispersion are poured and homogenized rapidly in 1 mL of the solvent to be studied, and simultaneously the scattered intensity $I_S$ and diameter $D_H$ are measured by light scattering. After mixing, the measurements are monitored over a 2-hour period, and subsequent measurements are made after 24 hours and 1 week. Nanoparticles are considered to be stable if their hydrodynamic diameter $D_H$ in a given solvent remains constant as a function of the time and equal to its initial value (given in Table III). Solvents surveyed here are DI-water (pH 7.4), phosphate buffer saline (PBS) and cell culture medium (Dulbecco's Modified Eagle's Medium, DMEM) with or without calf bovine serum.

Stability assays were performed on bare and coated iron oxide nanoparticles (Fig. 2). The commercial contrast agent Cliavist® was used as a benchmark. Bare particles with diameters 6.8 nm and 13.2 nm are stable at pH 1.8 (the pH at which they are synthesized) and stabilized by surface charges. Under rapid dilution, the pH variation induces a modification of the electrostatic charges at their surfaces. At physiological pH, the dispersions are close to their isoelectric point[54] (IP 6.7) and the particles are slightly positive. Bare particles hence aggregate due to van der Waals interactions. Citrate and carboxydextran (Cliavist®) coated particles are stable in PBS buffer. In culture media however, citrate coated particle precipitation occurs (Fig. 2) and it is attributed to the displacement of the ligands from the surfaces, as they are preferentially complexed by $Ca^{2+}$ and $Mg^{2+}$ divalent counterions.[9] For Cliavist®, the destabilization also takes place and come from the loosely tethered carboxydextran chains at the iron oxide surface. The particles coated with polymers, either poly(acrylic acid) or phosphonic acid PEG copolymers exhibit in contrast an outstanding stability over time. The dispersions are found to remain stable for months. For particles with polymer coats, protein corona was also not detected.[12,55] The stability results obtained in DMEM after one week are summarized in Fig. 2.

**RSC Advances**

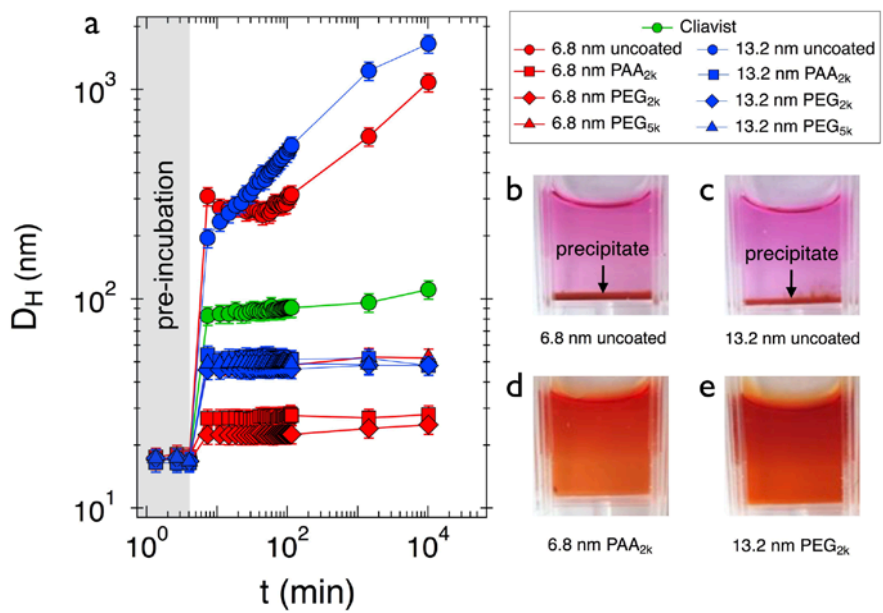

*Figure 2:* a) Hydrodynamic diameter $D_H$ versus time for contrast agents after they have been diluted in Dulbecco's Modified Eagle's Medium (DMEM). For uncoated particles, and for particles coated with citrates, aggregation and precipitation occur rapidly, as indicated by the initial size increase. For cliavist®, the aggregation is slower, but is still visible on the time scale of a week. Particles coated with poly(acrylic acid) and phosphonic acid PEG copolymers were found to remain stable for months. b-e) Images of some selected iron oxide dispersions in DMEM after one week. Stability assays were performed at experiments at the weight concentration of 0.1 wt. %.

## 3.3 - Relaxometry

The longitudinal and transverse relaxivities $r_1$ and $r_2$ of the coated particles dispersed in DI-water and in cell culture medium were measured using $T_1$- and $T_2$-weighted MR images on a 7.0 T on a 300 MHz micro-imaging Bruker spectrometer. Dispersion weighted phantom images at molar concentration $[Fe]$ = 0, 0.02, 0.05, 0.1, 0.2 and 0.5 mM were acquired, and their greyscale intensity was found to decrease with concentration, a result that is explained by the dipolar coupling between the magnetic moments and the water protons. In this range, the inverse of the relaxation times was found to vary linearly with $[Fe]$, according to: $1/T_{1,2} = r_{1,2}[Fe] + 1/T^0_{1,2}$ where $T^0_{1,2}$ denotes the water longitudinal and transverse relaxation time.[46,56] Fig. 3a and 3b display the transverse relaxivity $r_2$ histograms for the two particle sizes and for the various coatings. Comparison with Cliavist® is also shown (Fig. 3c). The relaxivities $r_{1,2}$ and relaxivity ratios $r_2/r_1$ are listed in the Supporting Information Section (S7).

Data in Fig. 3 show that the particle core size has a strong impact on the contrast, as $r_2$ passes from 62 mM$^{-1}$ s$^{-1}$ to 171 mM$^{-1}$ s$^{-1}$ for uncoated particles, and from 74 mM$^{-1}$ s$^{-1}$ to 226 mM$^{-1}$ s$^{-1}$ for the PEGylated ones. This increase is in agreement with that reported in Vuong et al.[42] who suggested that the iron based contrast agent transverse relaxivity grows quadratically with particle size. As for the coating, the data also show that the presence of a layer around particles leaves $r_1$ rather unchanged and increases $r_2$ by 20% in average. With respect to the solvent, the effect is the strongest with uncoated particles. From DI-water to cell culture medium, the transverse relaxivity of 6.8 nm $\gamma$-Fe$_2$O$_3$ is multiplied by a factor of 4 (from 62 mM$^{-1}$ s$^{-1}$ to 284 mM$^{-1}$ s$^{-1}$), whereas it is increased by a factor of 1.5 for the 13.2 nm particles (from 171 to 272 mM$^{-1}$ s$^{-1}$). This increase is consistent with particles being incorporated into large micron-sized aggregates, as shown in Fig. 2.[42,57] In conclusion, we have found that for maghemite the coating has a relatively weak impact on $r_1$ and $r_2$. These findings are important for the later *in vivo* MR imaging results, as they suggest that the role of the coating can be determined independently of other parameters.

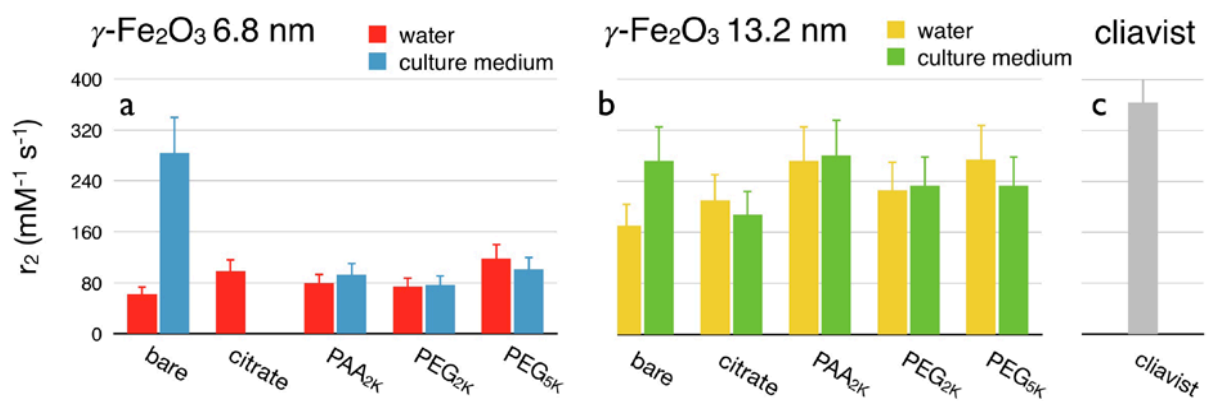

**Figure 3:** *Transverse relaxivity $r_2$ measured at 7 Teslas for 6.8 nm (a) and 13.2 nm (b) iron oxide nanoparticles dispersed in aqueous media. The data for Cliavist® (c) are shown for comparison. The error bars are defined as the standard deviations.*

## 3.4 - Biocompatibility and toxicity

Murine hepatocyte cells BWTG3 viability was studied using the colorimetric assay WST-1. WST-1 was performed at iron dose levels between $10^{-3}$ mM to 10 mM, to cover a broad concentration range and specifically the concentration used *in vivo* ($[Fe]$ = 0.5 mM in the blood pool). Fig. S8 a-f display the percentage of viable BWTG3 cells treated during 24 h with Cliavist® and with the 6.8 nm and 13.2 nm iron oxides. For the 11 particles studied, the viability exhibits a slight to moderate decrease around $[Fe]$ = 0.5 - 5 mM, the effect being the strongest for bare and citrate coated particles. This latter result could be related to the particle agglomeration and sedimentation, which increase the amount of materials adsorbed at the plasma membrane.[55] Observations of cells *in situ* using optical microscopy indicate that the cellular morphology was modified at high dose levels. Except for bare particles, the BWTG3 hepatoma cells exhibit a viability above 50% for all coating. These findings corroborate results obtained with other cell lines, including NIH/3T3 mouse fibroblasts,[50] 2139 Human lymphoblasts[9] and RAW264.7 macrophages.[12] Viability experiments with the coating agents alone, including phosphonic acid PEG copolymers, citrate and PAA polymers, were also done and reveal no cytotocity at relevant concentrations.[9,12,50]

## 3.5 - $T_2$-weighted MR Imaging of the liver: particle size and coating effects

The uptake of intravenously injected contrast agents was monitored by measuring the signal enhancement on $T_2$ weighted images of different organs. The iron oxide dispersion injected to the mouse was 100 μL and the iron concentration 5 mM (corresponding to 16.7 μmol kg$^{-1}$ or 0.93 mg kg$^{-1}$). Previous reports have shown that iron oxide based contrast agents exhibit uptake times of the order of minutes and clearance times in the range of days.[21-23,58-60] A logarithmic temporal scale was hence used for imaging the mouse organs at times ranging from one minute to one week. Fig. 4 compares MR scans obtained in the first three hours for Cliavist® (Fig. 4a to d) and for the 6.8 nm PEG$_{2K}$ particles (Fig. 4e to h). With the commercial agent, the mouse liver section exhibits a negative contrast enhancement shortly after injection (e.g. in Fig. 4b at 30 min). The darkening in the liver arises from the iron accumulation in the organ.[1,21,22,58,60-63] Multi-phosphonic acid PEG coated particles display a different behavior (Fig. 4e-g). For this sample the darkening is delayed by about 2 – 3 hours as compared to that of Cliavist®. During the period, the MR images show no contrast change of the liver and of other organs (spleen, kidneys), indicating a prolonged circulation in the bloodstream, as well as a delayed hepatic uptake (S6). At 3 hours post-injection (Fig. 4d and 4h), MR images exhibit a negative contrast enhancement as with cliavist®. A movie comparing the liver MRI intensity after injection of $PAA_{2K}$ and multi-phophonic acid $PEG_{2K}$ coated particles has been added as Supporting Information.

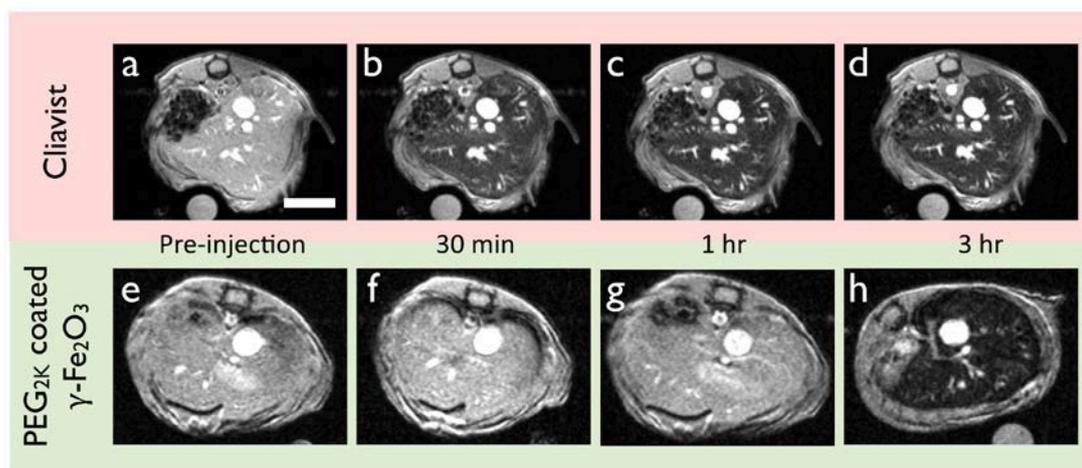

*Figure 4*: Magnetic resonance imaging scans for wild-type female 8 weeks BALB/c mouse livers at different time points (30 min, 1 and 3 hours) after injection for Cliavist® (a to d) and for the 6.8 nm $PEG_{2K}$ coated particles (e to h). In the assays, 100 µL of a $[Fe]$ = 5 mM dispersion were injected to the mouse. The corresponding dose is 16.7 µmoles, or equivalently 0.93 mg of iron per kilogram of mouse. A comparison of the two time sequences shows a delay in the negative contrast enhancement at about 3 hours for the PEG particles, compared to that of the commercial contrast agent.

To quantify the MR contrast, the liver mapping was performed by defining manually regions of interest (ROI), and by integrating the grey scale intensity as an 8-bit integer (between 0 and 255). Spatially average intensities were obtained and normalized with respect to the pre-injection level. With these conventions, the MRI signal intensity varies between 1 and 0, and the liver darkening is associated with a decrease as a function of the time (Fig. 5). The procedure was carried out at each time point for 8 particles: cliavist® ($n$ = 3), 6.8 nm particles coated with $PAA_{2K}$ ($n$ = 3), $PEG_{2K}$ ($n$ = 4) and $PEG_{5K}$ ($n$ = 2), and with 13.2 nm coated with citrate ($n$ = 1), $PAA_{2K}$ ($n$ = 2), $PEG_{2K}$ ($n$ = 2) and $PEG_{5K}$ ($n$ = 1). Here, $n$ designates the number of mice tested. Intravenous injection into the tail vein was performed as a single bolus in the conditions of Fig. 4, *i.e.* 100 µL of dispersion at 5 mM. Fig. 5 illustrates the correspondence between the liver MR images and the greyscale intensity. In the pre-injection period, the intensity remains at the level of 1. The MR signal then decreases rapidly after injection and passes through a minimum.

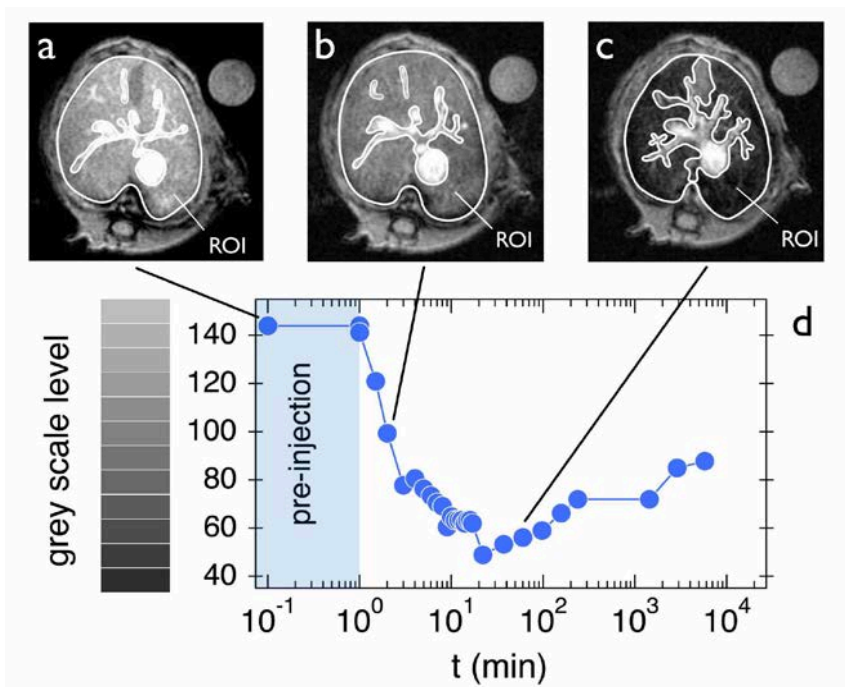

*Figure 5*: a) to c) Magnetic resonance imaging scans of a wild-type female 8 week BALB/c mouse liver at different time points after injection for the 13.2 nm poly(acrylic acid) coated particles. The liver mapping (indicated as ROI in the images) was performed manually by defining regions of interest, and by integrating the greyscale intensity as an 8-bit integer. d) Illustration of the correspondence between the MR images and the greyscale intensity displayed in the left-hand side panel. In the pre-injection period, the intensity remains constant and then decreases rapidly after injection.

In Fig. 6a and 6b, data for 6.8 nm and 13.2 nm particles coated with $PAA_{2K}$ and $PEG_{2K}$ are compared. For the poly(acrylic acid) coated nanoparticles, as well as for Cliavist®, the MRI signal intensity exhibits a broad minimum down to 0.35. For these samples, the minimum extends over a period of tens to hundreds of minutes. At longer timescale, typically from 1 to 7 days after injection the MRI signal intensity increases again and reaches unity, indicating that the liver returns to its pre-injection level. This recovery is associated with the iron clearance out of the liver.[21,58-60] On the opposite, PEGylated 6.8 nm and 13.2 nm particles exhibit prolonged circulations in the blood pool, the contrast enhancement appearing approximately after 2 – 3 hours. Later, the MRI signal passes through a minimum as in the previous case, and recovers its pre-injection limit after a few days. Interestingly, the MRI level at the minimum (around 0.6) remains higher than for anionic particles, indicating a lower iron oxide concentration and a reduced uptake

Concerning the polymer molecular weight, earlier studies have shown that the coating efficiency increases with the layer thickness.[1-3,6,15,23,25,26] To test this assumption, the pharmacokinetics of identical core particles with 2 kDa and 5 kDa PEG coating was investigated. From light scattering experiments, the layer thickness associated to $PEG_{2K}$ and $PEG_{5K}$ polymers was estimated at 5 nm and 8 nm, respectively (Table III). Fig. 7a and 7b compare the MRI signal intensity for the two particles. For 6.8 nm particles, the hepatic uptake is progressive and slightly more rapid for $PEG_{5K}$ than for $PEG_{2K}$. The difference may come from the hydrodynamic diameter that is about twice larger with the longer polymer, or from a lower PEG density (S5). For the 13.2 nm particles, the time dependent contrast enhancements remain similar in terms of hyposignal kinetics and amplitude. Combined together, these results show that the optimum stealthiness *in vivo* is obtained for the 6.8 nm particles and a 2 kDa PEG coating.

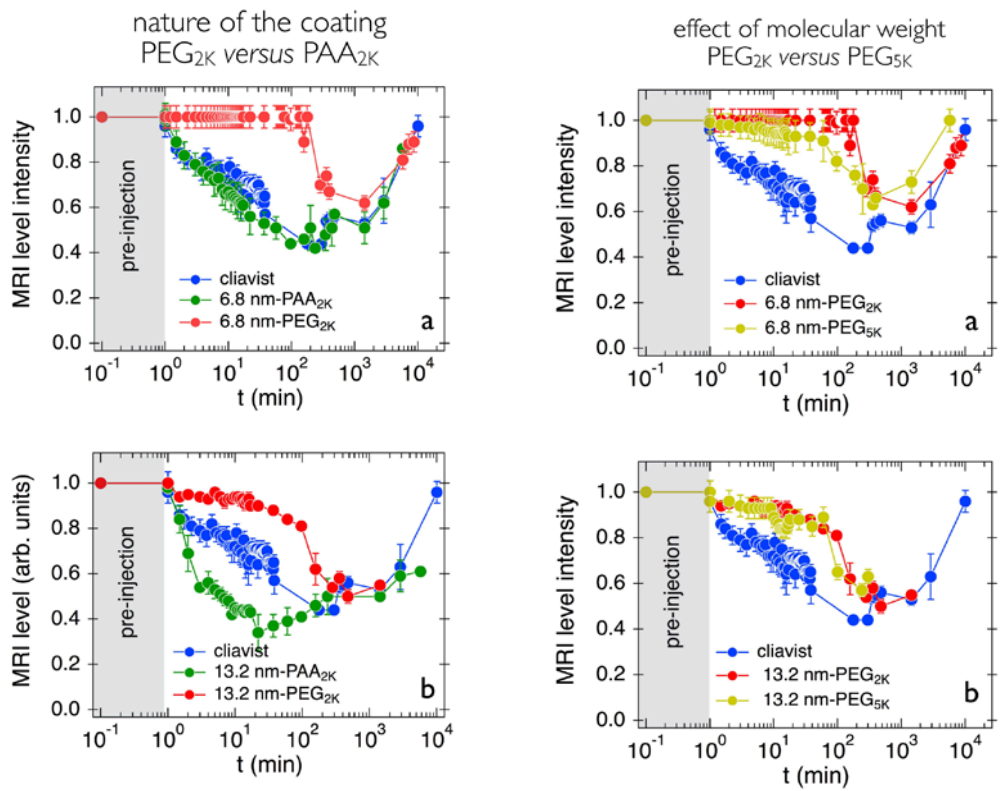

**Figure 6**: *Time dependences of the MRI signal intensity for a) 6.8 nm and b) 13.2 nm PAA$_{2K}$ and PEG$_{2K}$ coated nanoparticles. Cliavist® is shown for comparison. In the in vivo experiments, the iron oxide dispersion injected to the mouse was 100 µL and the iron concentration 5 mM (corresponding to 16.7 µmol kg$^{-1}$ or 0.93 mg kg$^{-1}$). With the conventions used, the darkening of the liver is associated to a decrease of the MRI signal as a function of the time. The error bars indicate the standard deviations.*

**Figure 7**: *Time dependences of the MRI signal intensity for a) 6.8 nm and b) 13.2 nm nanoparticles coated with PEG$_{2K}$ and PEG$_{5K}$. Cliavist® is shown for comparison. The experimental conditions are similar to those of Fig. 6. The error bars indicate the standard deviations.*

## 3.6 - Quantitative analysis of iron uptake and release by the liver

To retrieve the contrast agent pharmacokinetics parameters, the MRI signal intensity was adjusted using an expression of the form:

$$F_{MRI}(t) = \alpha + (1-\alpha)exp\left[-\left(\frac{t}{\tau_{up}}\right)^{\beta}\right] + (1-\alpha)\left(1 - exp\left[-\left(\frac{t}{\tau_{clear}}\right)^{\gamma}\right]\right) \quad (2)$$

where $\alpha$ is the intensity minimum, $\tau_{up}$ and $\tau_{clear}$ the characteristic times for uptake and clearance. In the above model, it is assumed that the two mechanisms are occurring one after the other. In a first attempt to adjust the MRI intensities, single exponentials were used but were found to be inappropriate. The single exponentials were replaced by stretched exponentials characterized by exponents $\beta$ for the uptake and $\gamma$ for the clearance. Stretched exponential functions have been found to describe many phenomena in Nature, in particular in complex systems with broad relaxation time distribution.[64,65] In Eq. 2, $\tau_{up}$ and $\tau_{clear}$ are related respectively to the half-life times $t_{1/2}^{up}$ and $t_{1/2}^{clear}$ that characterize the initial and terminal processes, via the expressions $t_{1/2}^{up} = \tau_{up}(ln2)^{1/\beta}$ and $t_{1/2}^{clear} = \tau_{clear}(ln2)^{1/\gamma}$. In the following, the data will be discussed in terms of the half-life times to ease comparison with earlier work.[23,58] Fig. 8a–h show the MRI signal intensities obtained for the different systems and the least-square calculations results using Eq. 2 (continuous lines). For the eight samples considered, Eq. 2 provides a satisfactory outcome over the entire time range. The adjustable parameters $\alpha$, the half-life times $t_{1/2}^{up}$ and $t_{1/2}^{clear}$, and the stretched exponents $\beta$ and $\gamma$ are listed in Table IV.

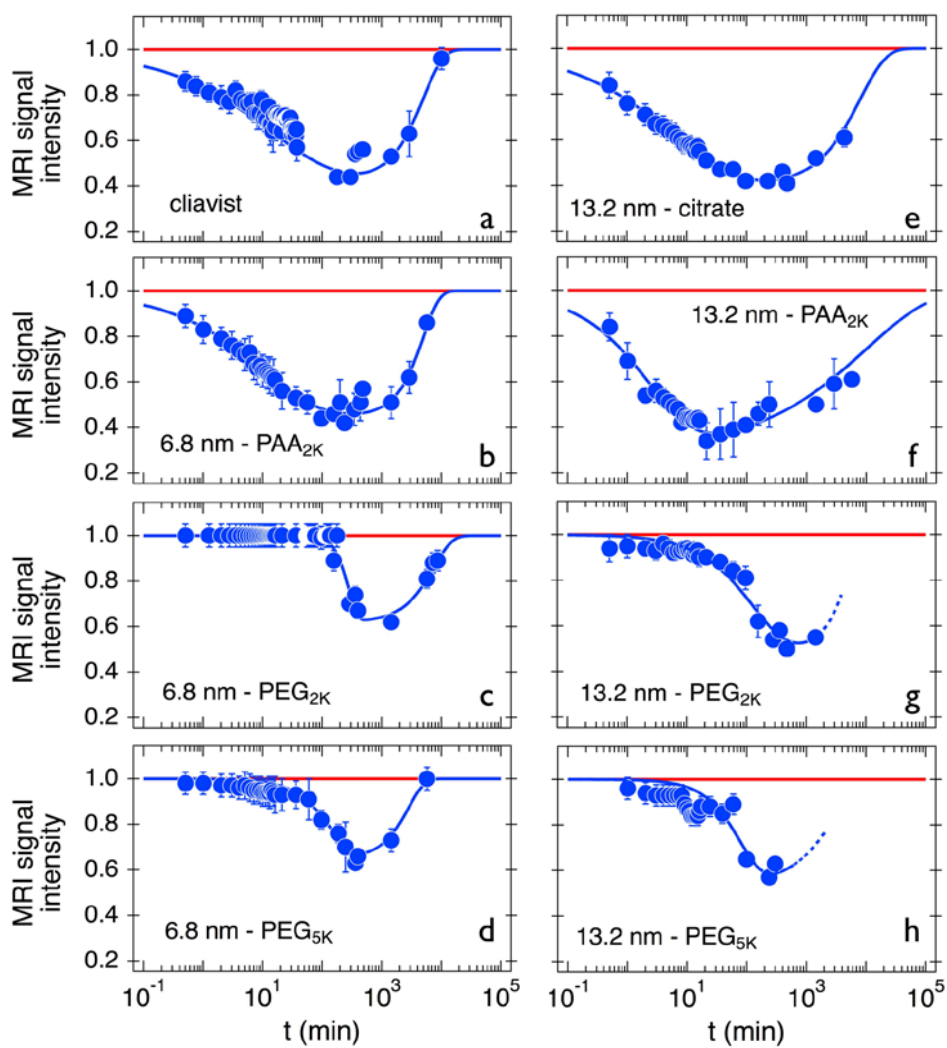

*Figure 8:* MRI signal intensity for the iron based contrast agents studied in this work, together with least-square calculations using Eq. 2 (continuous lines). The fitting of the uptake and clearance kinetics allows to determine the following parameters: the intensity minimum $\alpha$, the half-life times $t_{1/2}^{up}$ and $t_{1/2}^{clear}$ and the stretched exponents $\beta$ and $\gamma$ (Table IV).

As anticipated, the half-life time $t_{1/2}^{up}$ for the uptake mechanism depends strongly on the coating. For PAA$_{2K}$ 6.8 nm particles, the uptake time is estimated at 4.5 ± 0.3 min, whereas it is 55 times longer for PEG$_{2K}$ particles, at 249 ± 18 min. The stretch exponent for this mechanism varies from $\beta = 0.4 - 0.6$ for the agents that are rapidly uptaken, to $\beta = 3$ for the stealthiest ones. In Table IV, the size effect can be clearly seen as 6.8 nm particles have systematically prolonged circulation times as compared to the 13.2 nm particles.[1] The results suggest that the uptake mechanism depends primarily on two parameters, the probe size and the coating type. In the present study, the effects of each of these parameters are separately identified.

The clearance phenomenon is associated with the particle degradation and with the iron incorporation into the main intracellular iron storage proteins (ferritin, hemosiderin).[58,59] The MRI signal intensity returns to its pre-injection level as iron oxide particles are metabolized and their degradation products cleared (Fig. 8). As can be seen from Table IV, a rather uniform kinetics is observed for the clearance, with half-life times between 2 to 4 days. Indeed, $t_{1/2}^{clear}$ does not depend on the magnetic core sizes, nor on the coating. The benchmark agent Cliavist® exhibits also a comparable clearance time ($t_{1/2}^{clear}$ = 2.6 ± 0.1 day).

| $\gamma$-Fe$_2$O$_3$ | coating | $\alpha$ | $t_{1/2}^{up}$ (min) | $\beta$ | $t_{1/2}^{clear}$ (day) | $\gamma$ |
|---|---|---|---|---|---|---|
| 6.8 nm | PAA$_{2K}$ | 0.45 | 4.5 ± 1.3 | 0.46 | 2.8 ± 0.2 | 1.6 |
| | PEG$_{2K}$ | 0.62 | 249 ± 18 | 3.0 | 4.0 ± 0.3 | 1.4 |
| | PEG$_{5K}$ | 0.65 | 103 ± 15 | 1.2 | 1.5 ± 0.3 | 1.6 |
| 13.2 nm | citrate | 0.40 | 2.5 ± 0.2 | 0.42 | 3.9 ± 0.2 | 1.0 |
| | PAA$_{2K}$ | 0.30 | 1.4 ± 0.3 | 0.60 | 2.8 ± 0.3 | 0.4 |
| | PEG$_{2K}$ | 0.50 | 82 ± 13 | 0.80 | 2.8 ± 0.6 | 1.8 |
| | PEG$_{5K}$ | 0.54 | 59 ± 15 | 1.2 | n.d. | n.d. |
| Cliavist® | Carboxy dextran | 0.42 | 6.0 ± 1.2 | 0.40 | 2.6 ± 0.1 | 1.3 |

*Table IV*: Parameters obtained from the adjustment of the MRI signal intensity with Eq. 2. $\alpha$ denotes the intensity minimum, $t_{1/2}^{up}$ and $t_{1/2}^{clear}$ the half-life times associated to the uptake and clearance mechanisms and $\beta$ and $\gamma$ the stretched exponential exponents. The error bars are defined as the standard deviations (SD).

# 4 - Conclusion

In this work, novel iron oxide based contrast agents dedicated to *in vivo* magnetic resonance imaging were synthesized, with the aim of understanding the relationship between the particle coating properties and their pharmacokinetics. Maghemite ($\gamma$-Fe$_2$O$_3$) particles with diameters 6.8 nm and 13.2 nm were prepared and coated following bottom-up assembly processes. Different organic macromolecules were considered and a total of 10 different probes were evaluated both *in vitro* and *in vivo*. Statistical copolymers (Specific Polymers®, France) combining multiple phosphonic acid groups and PEG chains on the same backbone were synthesized and used as coats for iron oxide particles. These polymers were characterized by 2 kDa and 5 kDa PEGylated side-chains.

The particles were first evaluated regarding their physico-chemical properties and characterized in terms of size, charge and coating thickness. The colloidal stability, protein adsorption in culture media, contrast agent relaxometry and probe toxicity were investigated. The first result that emerges from the *in vitro* assay is that polymers, either poly(acrylic acid) or poly(ethylene glycol) are efficient coatings, as they prevent particles to agglomerate in culture media or to be covered by plasma proteins. The particles and the PEG based polymers were found to be non-toxic and biocompatible at the *in vivo* dose level.

The dispersions were injected intravenously into the mouse tail vein at a relatively low dose (16.7 µmol of iron per kilogram of mouse) compared to earlier reports. Liver, spleen and kidneys $T_2$-MR images were acquired prior and after injection, at time points between one minute and seven days. As nanoparticles are eliminated by the reticulo-endothelial system and filtered by the liver, the change in the liver MRI contrast was investigated and provided a direct indication of the nanoparticle pharmacokinetics. With this protocol, the particle uptake and clearance time evolution in and out of the liver could be retrieved. The main outcome from the *in vivo* MRI mouse assays is that coating appears as the primary parameter that affects the liver uptake kinetics and elimination from the blood pool. Phosphonic acid $PEG_{2K}$-copolymers provides the most efficient contrast agent in terms of stealthiness, as they are able to prolong the blood particle lifetime up to 250 minutes. In similar conditions, the life times for citrate and $PAA_{2K}$ coated particles are typically 50 times shorter. Although poly(acrylic acid) is a robust coating ensuring colloidal stability *in vitro*, its *in vivo* performance is comparable to that of citrate or dextran. The time at which the MRI signal intensity returns to its pre-injection level was found to follow a more universal behavior. The clearance time was comprised between 2 to 4 days for all samples studied, including the commercial benchmark Cliavist®. In conclusion, we have shown that a 5 nm layer made of multi-phosphonic acid poly(ethylene glycol) at the iron oxide surface significantly improve the colloidal stability, protein adsorption properties and stealthiness *in vivo* of MRI probes. Our work also demonstrates that the phosphonic acid coupled to innovative polymer synthesis may open up new horizons regarding the surface functionalization.

## Supporting Information

The Supporting Information includes sections on the iron oxide nanoparticle characterization in S1 (transmission electron microscopy), S2 (electron beam microdiffraction), S3 (vibrating sample magnetometry) and S4 (zeta potential). The PEG density determination at the particle surface is described in S5, the nanoparticle distribution kinetics in abdominal aorta in S6, the longitudinal and transverse relaxivity measurements in S7 and the cytotoxicity studies using murine hepatocytes in S8.

## Acknowledgements

We thank Virginie Berthat, Nicoletta Giamblanco, Isabelle Margaill, Giovani Marletta, Fanny Mousseau, Olivier Sandre and Leticia Vitorazi for fruitful discussions. The Laboratoire Physico-chimie des Electrolytes, Colloïdes et Sciences Analytiques (UMR Université Pierre et Marie Curie-CNRS n° 7612) is acknowledged for providing us with the magnetic nanoparticles. ANR (Agence Nationale de la Recherche) and CGI (Commissariat à l'Investissement d'Avenir) are gratefully acknowledged for their financial support of this work through Labex SEAM (Science and Engineering for Advanced Materials and devices) ANR 11 LABX 086, ANR 11 IDEX 05 02.